\documentclass[12pt,preprint]{aastex}

\begin{document}


\title{Comparing the observational instability regions for pulsating pre-main sequence and classical $\delta$ Scuti stars}
\shorttitle{PMS and $\delta$ Scuti instability strips}


\author{K. Zwintz\altaffilmark{1}}
\affil{Institute of Astronomy, T\"urkenschanzstrasse 17, A-1180 Vienna, Austria}
\email{zwintz@astro.univie.ac.at}

%\altaffiltext{1}{Visiting Astronomer, Cerro Tololo Inter-American Observatory.
%CTIO is operated by AURA, Inc.\ under contract to the National Science
%\altaffiltext{2}{Society of Fellows, Harvard University.}
%\altaffiltext{3}{present address: Center for Astrophysics,
%    60 Garden Street, Cambridge, MA 02138}
%\altaffiltext{4}{Visiting Programmer, Space Telescope Science Institute}
%\altaffiltext{5}{Patron, Alonso's Bar and Grill}


\begin{abstract}
A comparison of the hot and cool boundaries of the classical instability strip with observations has been an important test for stellar structure and evolution models of post- and main sequence stars.
Over the last few years, the number of pulsating pre-main sequence (PMS) stars has increased significantly: 36 PMS pulsators and candidates are known as of June 2007. This number allows to investigate the location of the empirical PMS instability region and to compare its boundaries to those of the classical (post- and main sequence) instability strip. Due to the structural differences of PMS and (post-)main sequence stars, the frequency spacings for nonradial modes will be measurably different, thus challenging asteroseismology as a diagnostic tool.
\end{abstract}


%% subject header.  Objects should be in the appropriate "individual"
%% headers (e.g. quasars: individual, stars: individual, etc.) with the
%% additional provision that the total number of headers, including each
%% individual object, not exceed six.  The \objectname{} macro, and its
%% alias \object{}, is used to mark each object.  The macro takes the object
%% name as its primary argument.  This name will appear in the paper
%% and serve as the link's anchor in the electronic edition if the name
%% is recognized by the data centers.  The macro also takes an optional
%% argument in parentheses in cases where the data center identification
%% differs from what is to be printed in the paper.



\section{Introduction}

The pre-main sequence (PMS) evolutionary phase is the short time span between the birth of a star from interstellar clouds and the onset of hydrogen burning upon its arrival on the main sequence. Gravitational contraction is the main energy source during this stage.

Low-mass PMS stars (i.e., with masses lower than $\sim$1.5 $M_{\odot}$) that have recently become visible in the optical range and contract along their Hayashi tracks \citep{hay61}, presumably evolve to T Tauri stars \citep{joy42}. They primarily have spectral types ranging from late F to M.
Intermediate-mass young stars (i.e., with masses between $\sim$1.5 and 10$M_{\odot}$) evolve more rapidly to Herbig Ae/Be (HAEBE) objects \citep[e.g.,][]{her60,fin84} with spectral types B to early F. The evolutionary stage of a HAEBE star is ambiguous: stars with masses below $\sim$4 $M_{\odot}$ are still in their PMS phase, while hotter stars have already started nuclear hydrogen burning before they become visible in the optical range. Members of both groups, T Tauri and HAEBE stars, show regular and irregular photometric and spectroscopic variability on very different timescales, indicating that stellar activity begins in the earliest phases of stellar evolution.

A and early F type PMS stars with masses between 1.5 and 4$M_{\odot}$ have the right combination of effective temperature, luminosity and mass to become vibrationally unstable.
With the now considerably increased number of known PMS pulsators and candidates it is possible to compare the instability regions for pre- and (post-) main sequence stars.

\section{Pulsating pre-main sequence stars}

After the discovery of the first two pulsating PMS stars in the young cluster NGC 2264 \citep{bre72}, it took more than 20 years until another PMS pulsator was found.
\citet{kur95} observed the Herbig Ae field star HR\,5999 and detected $\delta$ Scuti-like pulsations with an amplitude of $\sim$\,13\,mmag in Johnson $V$ and a period of $\sim$\,5.0 hours in the presence of irregular 350\,mmag background variability coming from the obscuring material in which the star is embedded.
These measurements enabled for the first time the examination of the internal structure of a PMS star using asteroseismology and to put constraints on the pulsation models \citep{mar98}.
Since then, several detections of pulsations in Herbig Ae field stars \citep[e.g.,][]{don97,mar00}, as well as in members of young clusters \citep[e.g.,][]{zwi05,zwi06} have been published.

The PMS pulsators have the same spectral types and luminosities as the classical $\delta$ Scuti stars. Hence, it is expected that their pulsation is driven by similar mechanisms, i.e., the $\kappa$ and $\gamma$ mechanisms in the hydrogen and helium ionization zones \citep{mar98}.
The unstable modes in pulsating PMS stars known so far are similar to those for classical $\delta$ Scuti stars, namely low radial order p- and g-modes \citep{sur01}.
Compared to post-main sequence stars, the inner parts of PMS stars are more uniform in density and chemical composition and without the presence of nuclear reactions, which is the reason for a lack of avoided crossing \citep{sur01}.
Frequencies of $\ell = 0$ modes with same radial order are nearly identical for pre- and post main sequence stars \citep{sur01}. For nonradial modes ($\ell > 0$) the patterns are more complicated due to evolutionary changes in the stellar interior and allow a discrimination between PMS and (post-) main sequence stars \citep{gue07}. These are very interesting differences which can be tested with asteroseismology.
Using single-site CCD time series observations and the recently developed nonradial pulsation models for PMS stars, it was possible for the first time to identify the 5 significant pulsation frequencies in NGC 6383 27 as $\ell$ = 0, 1 and 2 modes \citep[][ their star number 170]{zwi07}.

Until recently, 36 pulsating PMS stars have been discovered, of which 30 are bona fide PMS pulsators and 6 remain pulsating PMS candidates because their periods could not be determined accurately enough yet.  Using dedicated time series photometry, PMS pulsation has been discovered in 18 members of 6 young open clusters and in 18 Herbig Ae field stars.
A complete catalog of the presently known pulsating PMS stars and candidates including an overview of their parameters is given in Table \ref{allpmspuls2006}. Consequently, the instability region for PMS pulsators can be investigated observationally and compared to the classical $\delta$ Scuti instability strip.

\section{Verification of PMS nature}
The evolutionary stage of a star with given effective temperature ($T_{\rm eff}$), luminosity, and mass may be ambiguous, as the evolutionary tracks for pre- and post-main sequence stars intersect several times \citep[e.g., see Figure 1 in][]{bre98}. Thus, the PMS nature of a star has to be assessed by other means than the position in the HR diagram.

\subsection{Cluster membership}
The six young open clusters that host PMS pulsators are all younger than 10 million years indicating that their A to F type members have not reached the ZAMS yet. Hence, a basic membership criterion was the location of the targets of interest in the respective cluster HR diagrams.

In addition, proper motion data from the TYCHO-2 \citep{hog00} and the ASCC2.5 Catalogues \citep[All-sky Compiled Catalogue of 2.5 million stars;][]{kha01,kha05} were used for a verification of their cluster membership, and, hence, their PMS nature. Unfortunately, for only three of the cluster stars (see Table \ref{pmsnature}) proper motion data are listed.

Only for the stars 53 and 38 in NGC 6530 membership probabilites are published to be 78\% and 68\%, respectively \citep{van72}.

For star NGC 6383 27, H$_{\alpha}$ is in emission, and a near infrared excess was detected by \citet{the85}.  These features are characteristic for HAEBE objects (see below), hence the PMS nature of NGC 6383 27 seems to be confirmed.

\subsection{Herbig Ae field stars}

Herbig Ae stars are known to be PMS objects because their observable characteristics (e.g., typical emission lines, infrared and ultraviolet excess) seem to originate from the protostellar material in which such young stars are still embedded.
Except for V 351 Ori and V 1247 Lac all pulsating PMS field stars (see Table \ref{allpmspuls2006}) are included in the Catalog of Herbig Ae/Be stars \citep{the94}.

\citet{vie03} included V 351 Ori (their star PDS 201) in their list of Herbig Ae/Be candidates and link this star with the Ori B star forming region. As the observed H$_{\alpha}$ emission is quite weak for this star, \citet{vie03} indicate that \makebox{V 351 Ori} might be already a more evolved Herbig Ae star.
Hence, its evolutionary stage is still disputed.

V 1247 Ori (PDS 192) shows symmetric H$_{\alpha}$ emission features without, or with only very shallow absorption features \citep{vie03}. It is -- like V 351 Ori -- suggested to belong to the Ori B star forming region and its PMS evolutionary phase is highly probable.

\section{Color transformation}
The observational instability strip for classical (post-) main sequence $\delta$ Scuti stars is usually given in absolute magnitude ($M_V$) versus dereddened Str\"omgren $(b-y)_0$ color \citep{rod01}.
For the pulsating PMS stars only limited additional information is available, both spectroscopic and photometric. While apparent magnitudes and Johnson $(B-V)$ are provided, Str\"omgren measurements exist for only 4 of the cluster and 7 of the field stars.
Hence, for a direct comparison of classical $\delta$ Scuti and pulsating PMS stars, $M_V$ and $(B-V)_0$ had to be derived for both star groups.
For the {\em cluster} stars the calculation of absolute magnitudes and dereddened colors was done using the cluster distances and color excess'  (see Table \ref{pmspulsinfo}).
The situation is not satisfactory for the 18 pulsating PMS {\em field} stars. Parallaxes are published for 16 of them, but of low quality.
For part of the stars, $M_V$ based on those parallaxes yields implausible values, i.e., the stars would be located below the ZAMS. The other parallaxes have large errors. For PDS 2, e.g., the error of the parallax is $\sim$3 times larger than the parallax itself (Table \ref{fpmspulsinfo}).

Furthermore the color excess $E(B-V)$ was not measured for any of the pulsating PMS field objects. Hence, the following general relation for reddening in the interstellar medium was applied \citep[e.g., ][]{voi91}:
\begin{equation}
\label{av}
A_V \sim 0.3 \, {\rm mag / kpc}\\
\end{equation}
\begin{equation}
A_V = 3 \cdot E(B-V)\\
\label{avcomp}
\end{equation}
where $A_V$ denotes the interstellar absorption in $V$. It has to be stressed that this can only be taken as a first estimate (see Table \ref{fpmspulsinfo}).

With the relations given in \citet{saao}, the $(b-y)_0$ colors of the classical $\delta$ Scuti stars were transformed into $(B-V)_0$ values. Any transformation is affected by errors, but they are smaller when transforming intermediate band to broad band photometry than vice versa. Consequently, we transformed the $(b-y)_0$ colors of the $\delta$ Scuti stars to $(B-V)_0$.

\section{Comparing PMS pulsators to classical $\delta$ Scuti stars}

\begin{figure}
\plotone{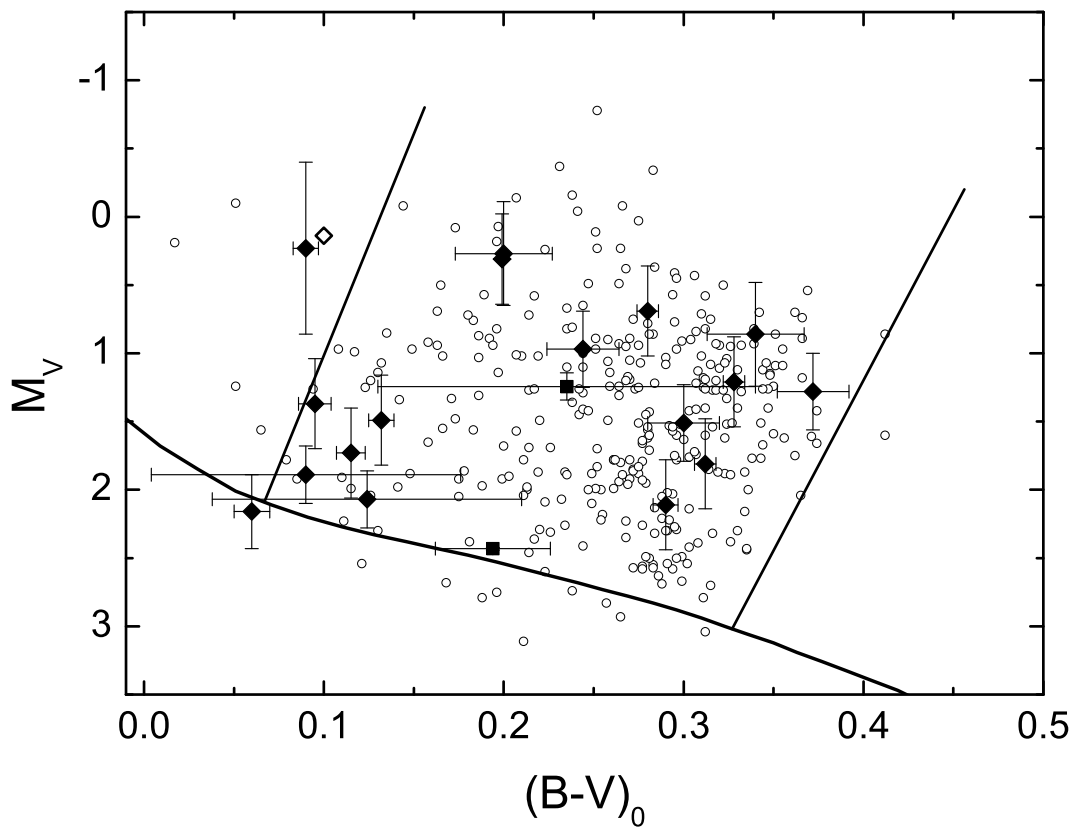}
\caption{Observational $M_V$ - $(B-V)_0$ HR diagram showing the ZAMS, the borders of the classical $\delta$ Scuti instability strip \citep[solid lines,][]{rod01}, the classical $\delta$ Scuti stars \citep[open circles,][]{rod01}, the two pulsating PMS field stars with sufficiently well known parallaxes (squares), and the 18 pulsating PMS cluster stars and candidates (diamonds). The open diamond marks the position of NGC 6823 230, for which no error bars could be determined.  \label{newstrip-BV}}
\end{figure}

It is now possible to compare classical $\delta$ Scuti to pulsating PMS stars in a common parameter space, $M_V$ vs. $(B-V)_0$, (Fig. \ref{newstrip-BV}).
For the $\delta$ Scuti stars $M_V$ is taken from \citet[][ open circles in Fig. \ref{newstrip-BV}]{rod01} and for the PMS pulsators and candidates $M_V$ was computed from the cluster distances (filled and open diamonds; Table \ref{pmspulsinfo}) and parallaxes of the field stars (filled squares; Table \ref{fpmspulsinfo}). 
For the majority of PMS field pulsators the published parallaxes either place the stars below the ZAMS or have large (i.e., $>$ 10\%) errors 
\mbox{(Table \ref{fpmspulsinfo})}. Hence, only two pulsating PMS field stars are selected for the HR diagram: $\beta$ Pic and HD 104237, with errors in the parallaxes of 1\% and 6\%, respectively.
According to \mbox{Figure \ref{newstrip-BV}}, PMS pulsators and $\delta$ Scuti stars seem to populate the same instability region in the HR diagram.
A lack of pulsating PMS stars is visible at the ``cool" corner of the instability region for classical $\delta$ Scuti stars. Whether this is only a selection effect caused by poor number statistics or has some astrophysical reason, can only be speculated about.

Figure \ref{newstrip-BV} also shows the uncertainties in $M_V$ and $(B-V)_0$ for the PMS pulsators. For the pulsating PMS {\em cluster} stars the errors in $M_V$ were computed from the errors in distance (Table \ref{pmspulsinfo}) and the $(B-V)_0$ errors were either taken from the literature, if available, or propagated from the listed errors in $V$ and $B$ measurements. In an analogous manner, the $M_V$ errors for the pulsating PMS {\em field} stars were calculated from the errors in the parallax (Table \ref{fpmspulsinfo}). 
As the parallax errors are typically larger than 10\%, only two of the 18 pulsating PMS field stars with accurate enough parallaxes could be used for the comparison of PMS pulsators and $\delta$ Scuti stars in the HR diagram.
For the pulsating PMS field stars errors for $(B-V)$ are scarce in the literature and we estimated them as standard deviations of independently measured and published $V$ and $B$ magnitudes. Note that the computed errors in $(B-V)_0$ for the field stars are at least 5 times larger than for all other stars. 

For NGC 6823 230 (open diamond) only a single value for $V$ and $B$ can be found in the literature without any error information. Hence, no errors in $(B-V)_0$ and also in $M_V$ can be determined for this star. It is remarkable, that - despite of the large formal errors - most PMS pulsators are located within the instability strip. This fact may indicate a too pessimistic error estimate for these stars.

Two outliers can clearly be identified: The PMS cluster stars, NGC 6823 230 (open diamond) and NGC 6823 279, are located bluewards of the blue border of the classical instability strip. The distance of NGC 6823 has the largest error compared to the other clusters containing pulsating PMS stars. The general cluster $E(B-V)$ of 0.845 mag adopted for the two PMS stars seems to be overestimated considering their position in a rather unobscured region of the cluster.

\section{Summary and Conclusions}

For 36 PMS stars ($\delta$ Scuti-like) pulsations were discovered by different authors within the last years. The PMS evolutionary phase of these stars has been assessed from observational evidences: either they are members of young open clusters or Herbig Ae stars.
Although only limited information is available in the literature for pulsating PMS stars, it is possible to explore their positions in the observational HR diagram. These positions and the observed period range support predictions of driving mechanisms for PMS pulsation that are similar to $\delta$ Scuti stars.
Furthermore, the hot and cool boundaries of the $\delta$ Scuti instability strip seem to coincide with the borders for the pulsating PMS stars.

It has to be noted that in general the determination of fundamental parameters for pulsating PMS stars is still unsatisfactory.
For only two of the 18 pulsating PMS field stars the parallaxes are accurate enough to be useful for a comparison to the $\delta$ Scuti stars.
Young clusters frequently exhibit strong differential reddening and field Herbig Ae stars are often surrounded by circumstellar gas and dust which severely limits the determination of $(B-V)_0$.
Temperature calibrations using $(B-V)$ colors exist for stars on the ZAMS \citep[e.g., ][]{ree98}, but their applicability to PMS stars that have luminosity classes III to V is questionable. Consequently, no $T_{\rm eff}$ and $L/L_{\odot}$ data are provided in this paper.

With a dense enough observed pulsation frequency spectrum it will be possible to discriminate directly between evolutionary stages using asteroseismology, as recently has been successfully demonstrated by \citet{gue07}. This avoids hazards with poorly determined distances and photometric indices

Whether the lack of PMS pulsators in the ``cool" corner of the classical instability strip has an astrophysical background or is due to poor number statistics, calls for dedicated observations.


%% Included in this acknowledgments section are examples of the
%% AASTeX hypertext markup commands. Use \url without the optional [HREF]
%% argument when you want to print the url directly in the text. Otherwise,
%% use either \url or \anchor, with the HREF as the first argument and the
%% text to be printed in the second.

\acknowledgments
Use was made of the WEBDA database (http://www.univie.ac.at/webda), developed by J.-C.Mermilliod (Laboratory of Astrophysics of the EPFL, Switzerland) and operated by E. Paunzen at the University of Vienna, Austria.
The author is very grateful to Werner W. Weiss for his constructive comments on the draft of this paper and his scientific and organizational efforts, which made this work possible. She also acknowledges funding through the Austrian Science Funds (FWF projects P-17580-N02 and T335-N16).
This work was enabled using the input of a number of colleagues, especially M. Breger, T. Kallinger, M. Marconi, A. Pamyatnykh, E. Paunzen and V. Ripepi.

\clearpage
\begin{deluxetable}{lcccrccr}
\tablewidth{0pt}
\tabletypesize{\scriptsize}
%\rotate
\tablecaption{Known pulsating PMS stars and candidates in clusters followed by the pulsating PMS field stars and candidates (as of June 2007) sorted by RA: names (for the cluster stars the numbering is according to the WEBDA database), equatorial coordinates (epoch 2000.0), spectral types (Sp), $V$ magnitudes, types of star (where `m' denotes a confirmed or very probable and `m?' a suspected cluster member and stars marked with `HAe' are known Herbig Ae type stars), numbers of detected pulsation frequencies (f), and the respective numbers of the references given below. \label{allpmspuls2006}}
\tablehead{
\colhead{Name} & \colhead{RA (2000.0)} & \colhead{DEC (2000.0)} & \colhead{Sp} & \colhead{$V$} &
\colhead{Type} & \colhead{f} &
\colhead{ref}\\
\colhead{ } & \colhead{[hh:mm:ss]} & \colhead{[dd:mm:ss]} & \colhead{ } & \colhead{[mag]} & \colhead{ } & \colhead{\#} & \colhead{ }
}
\startdata
IC 348 254 & 03:44:31.2 & +32:06:22.1 & F0/A8 III-IV & 10.60 & m & 4 & 1 \\
NGC 2264 2 & 06:39:05.9 & +09:41:03.4 & A7 III/IV & 9.73 & m & 12 & 2 \\
NGC 2264 20 & 06:39:28.5 & +09:42:04.1 & F2 III & 10.32 & m & 19 & 2 \\
NGC 6383 27 & 17:34:37.0 & -32:36:17.9 & A5 IIIp & 12.60 & m, HAe & 5 & 3 \\
NGC 6383 55 & 17:34:48.0 & -32:37:24.0 & -  & 12.83 & m? & 1 & 3 \\
NGC 6383 54 & 17:34:55.1 & -32:35:30.9 & F0 Ve & 12.34 & m? & suspected & 3 \\
NGC 6530 13 & 18:04:00.2 & -24:15:02.6 & -  & 13.35 & m? & 7 & 4 \\
NGC 6530 28 & 18:04:09.9 & -24:12:21.1 & - & 13.23 & m? & suspected & 4 \\
NGC 6530 38 & 18:04:13.9 & -24:13:28.0 & A0/A5 & 12.17 & m & 9 & 4 \\
NGC 6530 53 & 18:04:20.7 & -24:24:55.7 & A1 III & 13.07 & m & 5 & 4 \\
NGC 6530 57 & 18:04:21.8 & -24:15:46.9 & -  & 13.67 & m? & 1 & 4 \\
NGC 6530 78 & 18:04:30.8 & -24:23:42.1 & -  & 13.97 & m? & 3 & 4 \\
NGC 6530 159 & 18:04:42.3 & -24:18:03.5 & -  & 13.59 & m? & 2 & 4 \\
NGC 6823 230 & 19:43:06.8 & +23:16:37.8 & -  & 14.60 & m? & 2 & 5 \\
NGC 6823 279 & 19:43:09.1 & +23:17:49.6 & -  & 14.50 & m? & 2 & 5 \\
IC 4996 201 & 20:16:22.0 & +37:39:31.0 & A5 & 15.21 & m & 1 & 4 \\
IC 4996 171 & 20:16:30.0 & +37:39:32.8 & A4 & 15.03 & m & 1 & 4 \\
IC 4996 1085 & 20:16:43.9 & +37:42:26.5  & - & 15.30 & m & suspected & 4 \\
\tableline
PDS 2 & 01:17:43.5 & -52:33:30.8 & F2 & 10.73 & HAe & 3 & 6 \\
IP Per & 03:40:47.0 & +32:31:53.7 & A7 V & 10.40 & HAe & 9 & 7 \\
UX Ori & 05:04:30.0 & -03:47:14.3 & A3e & 9.60 & HAe & suspected & 8\\
HD 34282 & 05:16:00.5 & -09:48:35.4 & A0e & 9.85 & HAe & 2 & 9 \\
V 346 Ori & 05:24:42.8 & +01:43:48.3 & A5 III & 10.10 & HAe & 4 & 10 \\
HD 35929 & 05:27:42.8 & -08:19:38.4 & F0 IIIe & 8.20 & HAe & 1 & 11 \\
V 351 Ori & 05:44:18.8 & +00:08:40.4 & A7 IIIe & 8.90 & HAe & 5 &12 \\
CQ Tau & 05:35:58.0 & +24:44:54.0 & F2 IVe & 10.70 & HAe & 1 & 13 \\
BF Ori & 05:37:13.3 & -06:35:00.6 & A5 II-IIIevar & 10.30 & HAe & 1 (?) & 14 \\
V 1247 Ori & 05:38:05.3 & -01:15:21.7 & A5 III & 9.82 & HAe & 1 & 15 \\
$\beta$ Pic & 05:47:17.1 & -51:03:59.5 & A5 V & 3.86 & HAe & 2 (3?) & 16 \\
HD 104237 & 12:00:05.1 & -78:11:34.6 & A4 V & 6.60 & HAe & 2 (3?) & 17 \\
HD 142666 & 15:56:40.2 & -22:01:40.0 & A8 Ve & 8.81 & HAe & 1 & 18 \\
HR 5999 & 16:08:34.3 & -39:06:18.3 & A7 III/IVe & 6.98 & HAe & 1 & 19 \\
VV Ser & 18:28:49.0 & +00:08:39.0 & A2e & 11.50 & HAe & 2 (3?) & 20 \\
WW Vul & 19:25:58.7 & +21:12:31.0 & A3e & 10.51 & HAe & 1 (?) &  14 \\
PX Vul & 19:26:40.3 & +23:53:49.0 & F0 Ve & 11.67 & HAe & 1 (?) & 14 \\
V 375 Lac & 22:34:40.9 & +40:40:05.0 & A7e & 12.94 & HAe & 2 & 14 \\
\enddata
\tablerefs{
(1) \citet{rip02}; (2) \citet{bre72}; Kallinger \& Zwintz priv. comm.;
(3) \citet{zwi05}; (4) \citet{zwi06}; (5) \citet{pig00};
(6) \citet{ber06}; (7) \citet{rip06}; (8) Amado priv. comm.;
(9) \citet{ama04}; (10) \citet{pin03}; (11) \citet{mar00};
(12) e.g., \citet{rip03}; (13) Marconi \& Ripepi, priv. comm.; (14) \citet{ber05};
(15) \citet{lam90}; (16) \citet{koe03}; (17) \citet{kur99}; \citet{don97};
(18) \citet{kur01b}; (19) \citet{kur95}; (20) \citet{rip07};
}
\end{deluxetable}

\clearpage
\begin{table}
\begin{center}
\caption{Proper motion data from the TYCHO-2 and ASCC 2.5 Catalogues for three of the pulsating cluster stars. \label{pmsnature}}
\begin{tabular}{lrr}
\tableline\tableline
\multicolumn{1}{c}{Name} & \multicolumn{1}{c}{TYCHO-2 ${\mu}_{\alpha}$ / ${\mu}_{\delta}$} &
\multicolumn{1}{c}{ASCC 2.5 ${\mu}_{\alpha}$ / ${\mu}_{\delta}$} \\
 & \multicolumn{1}{c}{[mas/yr]} & \multicolumn{1}{c}{[mas/yr]} \\
\tableline
IC 348 254    & 10.0 / -6.5   & 7.73 / -5.99 \\
NGC 2264 2      & -6.1 / -5.7   & -6.11 / - 5.88 \\
NGC 2264 20     & - / - & -3.21 / -6.97 \\
\tableline
\end{tabular}
\end{center}
\end{table}

\clearpage
\begin{deluxetable}{lrccrcrc}
\tablewidth{0pt}
\tabletypesize{\scriptsize}
%\rotate
\tablecaption{Known pulsating PMS stars and candidates in clusters (as of June 2007): name, average distance and error, average distance modulus, the computed absolute magnitude $M_V$, $B-V$, $E(B-V)$, and $(b-y)_0$, where values marked with an asterisk were computed using the transformations given by \citet{saao}. \label{pmspulsinfo}}
\tablehead{
\colhead{Name} &  \colhead{dist}  & \colhead{$m-M$} & \colhead{$M_V$} & \colhead{$B-V$} &
\colhead{$E(B-V)$} & \colhead{$(b-y)_0$} & \colhead{Ref} \\
\colhead{ } & \colhead{[pc]} & \colhead{[mag]} & \colhead{[mag]} & \colhead{[mag]} &
\colhead{[mag]} & \colhead{[mag]} & \colhead{$B-V$ / $b-y$}
}
\startdata
IC 348 254    &  348$\pm$43 &        10.81    &   -0.21  &   0.97   &      0.93   &  $^{\star}$0.023 & 1 / -\\
NGC 2264 2   &    721$\pm$126 &          9.28   &   0.86   &   0.27   &      0.05   &   0.118 & 2 / 3  \\
NGC 2264 20   &    721$\pm$126 &    9.28   &   1.55   &   0.41   &      0.05   &   0.225 & 2 / 3 \\
NGC 6383 27    &  1200$\pm$157 &        11.9   &   1.00   &   0.70   &      0.33   &  $^{\star}$0.235 & 4 / - \\
NGC 6383 55    &  1200$\pm$157 &       11.9    &   0.93   &   0.63   &      0.33   &   $^{\star}$0.181 & 4 / - \\
NGC 6383 54    &  1200$\pm$157 &        11.9   &   0.44   &   0.57   &      0.33   &  $^{\star}$0.143 & 4 / -\\
NGC 6530 13    &  1537$\pm$237          &   11.65   &   1.70   &   0.45  &   0.33   & $^{\star}$0.062 & 5 / -\\
NGC 6530 28    &  1537$\pm$237      &   11.65   &   1.58   &   0.42  &   0.33   &  $^{\star}$0.041 & 5 / - \\
NGC 6530 38    &  1537$\pm$237          &   11.65   &   0.53   &   0.52  &   0.33   &  $^{\star}$0.111 & 5 / - \\
NGC 6530 53 &  1537$\pm$237    &        11.65   &   1.42   &   0.65   &      0.33   &  $^{\star}$0.195 & 5 / - \\
NGC 6530 57    &   1537$\pm$237         &    11.65   &   2.02   &   0.63  &   0.33   & $^{\star}$0.181 & 5 / - \\
NGC 6530 78 &  1537$\pm$237    &        11.65   &   2.32   &   0.61   &      0.33   &  $^{\star}$0.166 & 5 / - \\
NGC 6530 159  &  1537$\pm$237    &        11.65   &   1.94   &   0.44   &      0.33   &  $^{\star}$0.051 & 5 / - \\
NGC 6823 230   &  2377$\pm$688 &    14.01  &   0.17   &   0.93      &   0.85   &  0.068 & 6 / 7  \\
NGC 6823 279   &  2377$\pm$688 &   14.01   &   0.56   &   0.92     &   0.85   &  -0.073  & 8 / 7 \\
IC 4996 201  &  1651$\pm$155    &        13.28   &   1.93   &   0.80   &      0.67   &  $^{\star}$0.122 & 9 / -  \\
IC 4996 171  &  1651$\pm$155    &        13.28   &   1.75   &   0.75   &      0.67   &  $^{\star}$0.092 & 9 / -  \\
IC 4996 1085  &  1651$\pm$155    &        13.28   &   2.02   &   0.72   &      0.67   &  $^{\star}$0.103 & 9 / - \\
\enddata
\tablerefs{Numbering, distances, errors in distances, $m-M$ and $E(B-V)$ values are taken from the WEBDA database. The observed $(b-y)$ colors were dereddened using the complete set of $uvbyH_{\beta}$ measurements and with the {\tt TempLogg$^{\tt TNG}$} software \citep{templogg}. The colors are taken from (1) \citet{tru97}, (2) \citet{kwo83}, (3) \citet{han99}, (4) \citet{zwi05}, (5) \citet{sun00},
(6) \citet{hoa61}, (7) \citet{pen03}, (8) \citet{gue92}, (9) \citet{del98}.
}
\end{deluxetable}

\clearpage
\begin{deluxetable}{lrrcrccr}
\tablewidth{0pt}
\tabletypesize{\scriptsize}
%\rotate
\tablecaption{Known pulsating PMS field stars and candidates (as of June 2007): name, parallax, error in parallax, the computed absolute magnitude $M_V$, $B-V$ taken from the reference given in the next column, $E(B-V)$, and $(b-y)_0$, where values marked with an asterisk were calculated using the transformations given by \citet{saao}. \label{fpmspulsinfo}}
\tablehead{
\colhead{Name} & \colhead{par} & \colhead{e(par)}  & \colhead{$M_V$} &
\colhead{$(B-V)$}  & \colhead{Ref} & \colhead{$E(B-V)$} & \colhead{$(b-y)_0$}\\
\colhead{ } & \colhead{[mas]} & \colhead{[mas]}  & \colhead{[mag]} &
\colhead{[mag]} & \colhead{$(B-V)$} & \colhead{[mag]} & \colhead{[mag]}
}
\startdata
PDS 2 & 9.69 & 29.60 & 5.66 &   0.38  & 3  &  0.01 & $^{\star}$0.234 \\
IP Per  &   42.29 & 29.70  &    8.48   &   0.34 &  1  &     0.00   & $^{\star}$0.208  \\
UX Ori  &   0.69  & 2.47  &    -0.13  &   0.16  &  1 &     0.15   & 0.055  \\
HD 34282    &   5.89  & 1.62    &   3.68   &   0.33   &   1 &  0.02   &  $^{\star}$0.189 \\
V 346 Ori   &   1.39  & 2.01  &   0.97   &   0.20 &  1  &     0.07   &  $^{\star}$0.066  \\
HD 35929    &   0.66  & 0.93  &   -2.78  &   0.45 &  1  &   0.15   & 0.267  \\
V 351 Ori   &   3.55   & 1.59 &   1.66   &   0.38   &  1 &   0.03   &  0.120  \\
CQ Tau  &   10.00  &  2.00 &   5.60   &    0.50 &  1    &   0.01   & $^{\star}$0.319 \\
BF Ori  &   0.66  & 1.79  &   -0.78  &   0.25 &  1  &   0.15   & 0.129 \\
V 1247 Ori   &   32.4  & 25.10    &   7.38   &  0.34   &  1  &    0.00   &  $^{\star}$0.210  \\
$\beta$ Pic    &   52.00  & 0.50 &      2.43   &   0.20 &  1  &    0.00   &  0.085 \\
HD 104237   &   8.54  & 0.52  &   1.24   &   0.25 &  1  &   0.01   & 0.124  \\
HD 142666   &   3.29  & 10.89   &   1.40   &   0.56 &  1    &   0.03   &  $^{\star}$0.343  \\
HR 5999 &   4.88  & 0.87  &    0.53   &   0.31 &  1  &     0.02   & 0.154  \\
VV Ser  &   2.27 & -  &     3.65   &   0.93  &  2 &  0.90   &  $^{\star}$0.561  \\
WW Vul  &   47.7 & 26.89       &   8.99   &   0.38  &  1  &      0.00   &  $^{\star}$0.237  \\
PX Vul  &   -   &   - &       &   0.74  & 1 &    0.00   &  $^{\star}$0.458 \\
V 375 Lac   &   -   & - &     -   &   0.88   &   2 &   0.00   &  $^{\star}$0.527  \\
\enddata
\tablerefs{Parallaxes and their errors are taken from the Hipparcos Catalogue \citep{hip97} and the ASCC 2.5 Catalog \citep{kha01,kha05}; except for VV Ser for which the parallax was computed from the distance given in \citet{rip07}. The $(B-V)$ values were taken from (1) the ASCC 2.5 Catalog \citep{kha01,kha05}, (2) from the Third Catalog of Emission-Line Stars of the Orion Population \citep{her88} and (3) from \citet{vie03}. The measured $(b-y)_0$ values are taken from \citet{hau98}.}
\end{deluxetable}

%Figures

\clearpage

\end{document}